# Study of the Four-Rod RFQ Using the Normal Mode Theory of Transmission Lines

Valeri Kapin*
Moscow Engineering Physics Institute
Kashirskoe sh. 31, 115409, Moscow, Russia

*Abstract*

The normal mode theory of the transmission lines is applied to the studies of the Four-Rod RFQ. This resonator consists of four quadrupole electrodes and stems. It can be simulated by the four-conductor shielded transmission line, which is loaded by sets of coaxial transmission lines corresponding to the support stems. The ideal case of the indefinite periodic Four-Rod RFQ is considered. Formulas for resonance frequency, the voltage distribution along the electrodes and the shunt impedance are obtained.

## 1. INTRODUCTION

The accelerating structures as four-rod RFQ resonators are widely used for linear accelerators [1-5]. These resonators (fig.1 and fig.2) consist of four quadrupole electrodes (rods) 1 and 2, which are located in the tank 3 and connected with the tank by the stems 4.

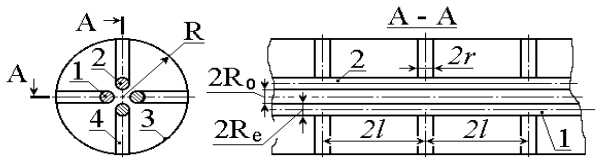

Fig.1. In-line stem 4-rod RFQ.

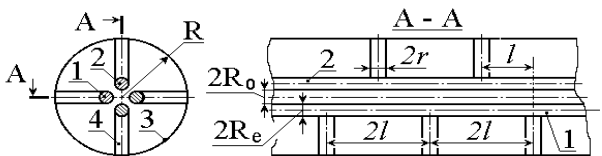

Fig.2. Alternate stem 4-rod RFQ.

In terms of the longitudinal connection points of the rods with the tank wall on the neighboring electrodes, all of these resonators can be grouped into two main types. The first type can be named as "In-line stem 4-rod RFQ"(fig.1). It is the way of connection of rods with the tank, where the connection points of all rods with the tank are located at the same longitudinal position. The second type can be named as "Alternate stem 4-rod RFQ". It is another way of connection of rods with the tank, where each set of opposing rods is connected to the tank at the middle of the successive supports of the other set of rods(fig.2).

In this paper, an equivalent circuit modeling of the 4-rod RFQ resonators is proposed. This approach can be considered as the extension of studies of 4-rod RFQ [2-5], which generally uses the conventional transmission line theory based on the concepts of distributed parameters and using a circuit-analysis approach. Here, operating *resonator* modes in 4-rod RFQ are studied using the theory of *propagating* modes (normal TEM modes).

## 2. THE MODEL OF FOUR RODS RFQ RESONATORS

Assuming only TEM waves propagating along all conductors of four rods RFQ resonator, this resonator can be simulated by the four-conductor shielded transmission line (4CSTL), which is loaded by sets of coaxial transmission lines (CTL's) corresponding to the stems. The equivalent circuits for both types of four rods RFQ resonators are presented in fig.3 and fig.4.

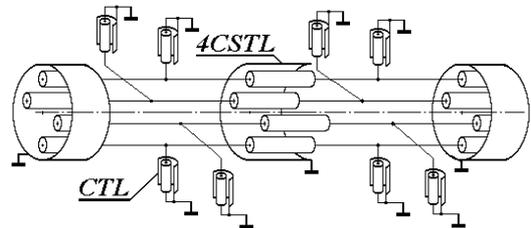

Fig.3. Equivalent circuit of "In-line stem 4-rod RFQ".

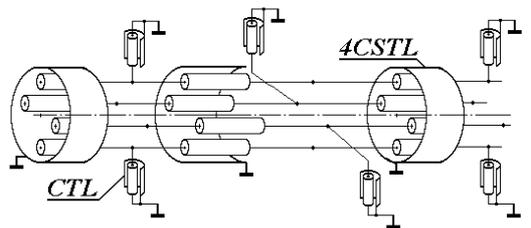

Fig.4. Equivalent circuit of "Alternate stem 4-rod RFQ".

If we consider the 4CSTL as a main part of these circuits, the difference between these two types of RFQ resonators is denoted by the ways of connection of the CTL's to the 4CSTL.

The propagation of TEM waves is described by the system of the well-known telegraph equations [6-8] for voltage and current on the conductors of the N-conductor transmission line, which are the complex functions of only the longitudinal coordinate $z$. This system consists of $2N$ differential equations and can be written in the matrix form:

$$\frac{d}{dz}\|U\| = -i\omega \cdot \|L\| \cdot \|I\|, \quad \frac{d}{dz}\|I\| = -i\omega \cdot \|\beta\| \cdot \|U\|, \quad (1)$$

where $\|U\|=[U_1,...,U_N]^T$ and $\|I\|=[I_1,...,I_N]^T$ are N-dimensional vectors of voltage and current, respectively; $\omega$ - is the circular frequency; $i^2 = -1$; $\|\beta\|$ and $\|L\|$ are N×N-dimensional square matrices of coefficients of the electrostatic induction and magnetostatic inductivity, respectively. The case of N=4 corresponds to the 4CSTL and N=1 to the CTL, while the shields are numbered as the (N+1)th conductors.

---

* now at Accelerator Lab., Inst. Chemical Research., Kyoto Univ.

## 3. NORMAL MODES FOR THE 4CSTL

In a transmission line with N conductors there are N normal TEM modes of propagation and any field is expressed by the superposition of these normal modes [7,8]. Mathematically the problem of a definition of normal modes is one of calculation of a so-called modal matrix $\|M\|$ [7,8], which allows conversion of the square matrices $\|\beta\|$ and $\|L\|$ of the initial system (1) into the diagonal form. It was found that the transformation to the normal modes can be expressed by:

$$\|\xi\| = \|D\| \cdot \|M\| \cdot \|U\|, \quad \|\eta\| = \|D\|^{-1} \cdot \|M\| \cdot \|I\|, \qquad (2)$$

$$\|M\| = \frac{1}{2} \cdot \begin{bmatrix} 1 & 1 & 1 & 1 \\ 1 & -1 & 1 & -1 \\ 1 & 1 & -1 & -1 \\ 1 & -1 & -1 & 1 \end{bmatrix}, \quad \|D\| = \begin{bmatrix} 1/2 & 0 & 0 & 0 \\ 0 & 1 & 0 & 0 \\ 0 & 0 & 1 & 0 \\ 0 & 0 & 0 & 1 \end{bmatrix},$$

where $\|\xi\|=[\xi_1,...,\xi_N]^T$ and $\|\eta\|=[\eta_1,...,\eta_N]^T$ are N-dimensional vectors of voltage and current of normal TEM modes, respectively; $\|M\|$ is the modal matrix for $\|\beta\|$ and $\|L\|$. $\|D\|$ is a normalizing matrix. The solution for normal modes is expressed as well as for the usual CTL:

$$\|\xi\| = \|A\| \cdot \cos(kz) + \|B\| \cdot \sin(kz),\qquad (3)$$
$$\|\eta\| = i\|Z\|^{-1} \cdot \{-\|A\| \cdot \sin(kz) + \|B\| \cdot \cos(kz)\},$$

where $k=\omega/v$ - the wave number; v - velocity of light; $\|A\|$ and $\|B\|$ are vectors of constant coefficients defined by boundary conditions; $\|Z\|$ and $\|C\|$ are diagonal matrices of wave impedance and capacitance per unit length for normal modes and $\|Z\|^{-1} = v \cdot \|C\|$.

Figure 5 illustrates normal TEM modes in the 4CSTL. The top figures show the equivalent circuits; the middle ones are the E-line patterns; the bottom ones are the two-dimensional electrostatic problem for definition of elements of $\|C\|$.

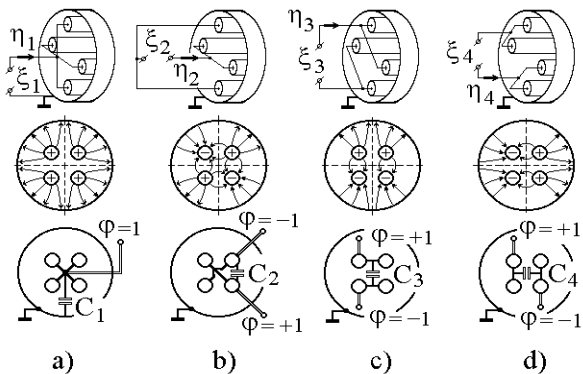

Fig.5. Normal TEM modes in the 4CSTL: the 1st, 2nd, 3rd and 4th modes corresponds to figures a), b), c) and d), respectively.

As can be seen, the mode fields have clear physical meaning. The first mode can be named as the coaxial, because it corresponds to the usual CTL with split four-wire central conductor. The second mode is a quadrupole one. The third and fourth modes are two dipole modes. The directions of their fields are perpendicular between each other. Obviously, only the second mode has suitable field for RFQ, that is, the other modes can be considered as parasitic ones. The coaxial mode does not penetrate the accelerating channel, while the dipole modes can distort the stable particle motion. Therefore, it is necessary to derive the conditions to suppress the dipole mode components in the operating resonator mode.

## 4. DIPOLE MODE COMPONENT SUPPRESSION

The sufficient conditions for suppressing dipole mode components are denoted by $\xi_3 \equiv 0$ and $\xi_4 \equiv 0$ at any $z$. For voltage and current on the 4CSTL conductors, this means that $U_1 \equiv U_3$, $U_2 \equiv U_4$, $I_1 \equiv I_3$ and $I_2 \equiv I_4$ at any $z$. From the above, necessary conditions for equivalent circuit parameters of stems can be obtained. According to the equivalent circuits in fig.3 and fig.4, stems are simulated by CTL's, therefore these necessary conditions can be formulated for electrical parameters of CTL's. They are expressed as the demands that the input impedance of CTL's connecting to opposing conductors of the 4CSTL should be the same. Figure 6 illustrates some configurations of stems for "Alternate stem 4-rod RFQ" and their equivalent circuits at assumption that stems are simulated by CTL's shorted at the end. The CTL's have the length $l_0$, wave impedance $Z_0$ and input impedance $Z_{in} = iX_0$, where $X_0 = Z_0 \text{tg}(kl_0)$. The cases a), b) and c) of the stems in fig.6 have the same equivalent stems impedance for opposing rods and hence, the necessary conditions are satisfied. In cases d) and e), these impedance are not equal due to some additional reactance $C_A$ and $L_A$, and the field will have dipole components. The proposed model allows the evaluation of the dipole fields with real stem configurations for studies of dipole component effects on beam dynamics.

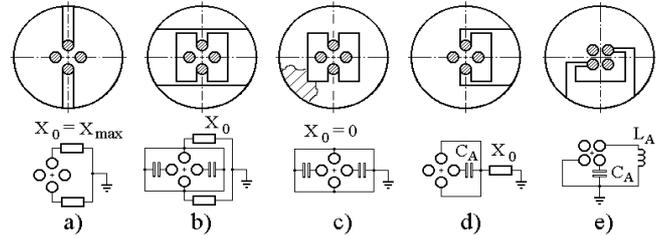

Fig.6. Real stem configurations and their equivalent circuits.

## 5. RF CHARACTERISTICS OF 4-ROD RFQ'S

RF characteristics of the two types of resonators (fig.1,2) were studied by equivalent circuits in fig.3 and fig.4. The calculations were carried out for two different configurations of stems, which have different values of $X_0$. The first one is shown in fig.6-a) (at $r=R_e=R_0$) and the second one in fig.6-c).

In the equivalent circuit of "In-line stem 4-rod RFQ" (fig.3), only quadrupole TEM mode component exists in the 4CSTL at the operating *resonator* mode. But in the equivalent circuit of "Alternate stem 4-rod RFQ" (fig.4), two TEM mode components (coaxial and quadrupole) exist in the 4CSTL at operating *resonator* mode. The following two equations are the resonance conditions for the first and second types of 4-rod RFQ, respectively:

$$4 \cdot X_0 \cdot \text{tg}(kl) = Z_2, \quad \text{tg}\left(\frac{kl}{2}\right) - \frac{Z_2}{4Z_1} \cdot \text{ctg}\left(\frac{kl}{2}\right) = -\frac{2X_0}{Z_1} \qquad (4)$$

Figure 7 shows $l/\lambda$ as functions of $R/\lambda$ at $R_0 = R_e = 1$cm.

2192

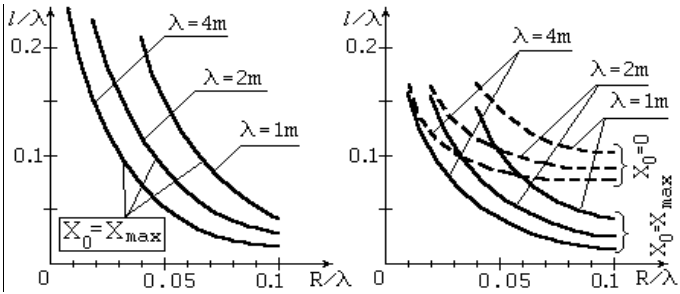

Fig.7. Resonance conditions for the 1st (left graph) and the 2nd (right graph) type of 4-rod RFQ.

The voltage flatness $\delta$ is defined as a maximum value of relative deviation of the voltage distribution $U_v(z)$ from its average value $\hat{U}_v$. For our model, $\delta$ is derived from $\xi_2(z)$-function and is defined by $\delta=(1-\chi)/(1+\chi)$, where $\chi=\cos(kl)$ for the first and $\chi=\cos(kl/2)$ for the second type of resonator, while the values $kl$ are defined by equations (4). $\delta$ are shown in fig.8 as functions of $R/\lambda$ at $R_0=R_e=1$cm.

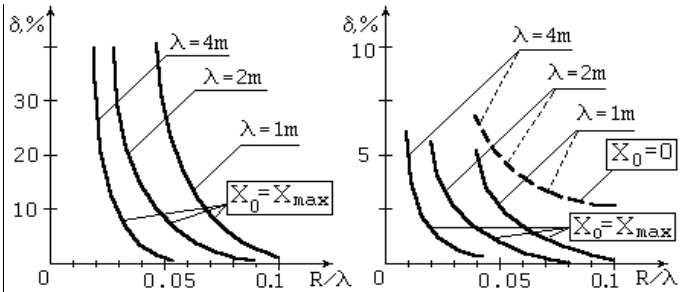

Fig.8. The flatness of voltage distribution for the 1st (left graph) and the 2nd (right graph) type of 4-rod RFQ.

The efficiency of RFQ structures is described by the specific shunt-impedance, $\rho$[1], defined by $\rho=\hat{U}_v^2/(2P/l)$, where P is the power loss, which is calculated from the distribution of the surface current density $j_S$ on the cavity surface. For our model, the longitudinal distribution of $j_S$ is defined by the solutions of the telegraph equations (1) and the transverse distribution of $j_S$ is obtained from the solution of the corresponding electrostatic problems for normal TEM modes (fig.5). Thus, this model takes into consideration the non-uniformity of current distribution on the conductors of the 4CSTL determined by so-called the proximity effect due to close spacing of the conductors [9]. The total power loss P is the sum of the loss $P_1$ on the 4CSTL and the loss $P_2$ on CTL's. For the 1st type of resonator these values are:

$$\frac{P_1}{l}=\frac{A_0 \cdot G_-(2l)}{\cos^2(kl)}\cdot[K_{2,R}+0.5\cdot K_{2,S}],\quad \frac{P_2}{l}=\frac{A_0\cdot B_0}{4l}\cdot\mathrm{tg}^2(kl),\quad (5)$$

$$A_0=\frac{R_S}{4\pi R_0}\cdot\left(\frac{4\chi}{1+\chi}\cdot v\cdot\hat{U}_v\cdot C_2\right)^2,\quad G_\pm(x)=\frac{1}{2}\cdot\left[1\pm\frac{\sin(kx)}{kx}\right],$$

$$B_0=\frac{R^*+(1/2k)\cdot\sin(2kR^*)}{\cos^2(kR^*)},\quad R^*=R-3R_e.$$

For the 2nd type of resonator, the power losses are:

$$\frac{P_1}{l}=\frac{A_0}{\cos^2(kl)}\cdot\{G_+(l)\cdot\mathrm{tg}^2(kl/2)[K_{1,R}+\frac{4R_0}{R}\cdot K_{1,S}]+$$
$$+G_-(l)\cdot[K_{2,R}+0.5\cdot K_{2,S}]\},\quad (6)$$

$$\frac{P_2}{l}=\begin{cases}A_0\cdot(8R_0/l)\cdot\mathrm{tg}^2(kl/2)\cdot\ln(R/2R_0),\text{ for }X_0=0\\ A_0\cdot(4B_0/l)\cdot\mathrm{tg}^2(kl/2),\text{ for }X_0=X_{max},\end{cases}$$

where $R_S$ - the surface skin resistance; the coefficients $K_{m,n}$, ($m$=1 or 2; $n$=R or S)- characterize the proximity effect on the conductors of the 4CSTL, while the first subscripts correspond to number of TEM modes and the second subscripts define a surface (S - shield; R - rods). Calculations showed that $K_{1,R}\approx 1.6$, $K_{1,S}\approx 1.0$, $K_{2,R}\approx 1.2$, $K_{2,S}\approx 0.0$. Figure 9 shows the calculated $\rho$ as functions of $R/\lambda$ at $R_0=R_e=1$cm for copper resonators ($R_S=4.5\cdot 10^{-3}\cdot\lambda^{-1/2}$, $\Omega$m).

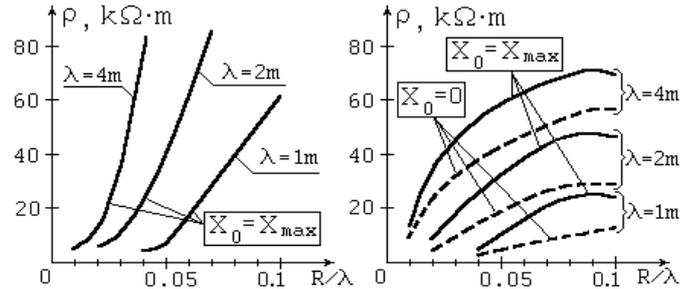

Fig.9. The specific shunt-impedance for the 1st (left graph) and the 2nd (right graph) type of 4-rod RFQ.

## 6. CONCLUSION

The presented equivalent circuit modeling allows the calculation of the RF characteristics of 4-Rod RFQ resonators for a wide range of geometrical parameters. It demonstrates the existence of an essential difference between "In-line stem 4-rod RFQ" and "Alternate stem 4-rod RFQ". From graphs it is seen that the first type of 4-rod RFQ has a better RF characteristics than the second one for large values of $R/\lambda$, but for small values of $R/\lambda$ it has a unsatisfactory voltage flatness. That is, there is some value of $R/\lambda$, at lower side of which the second type of resonator is preferable to the first one. Practically this means that if a tank diameter is fixed, the "Alternate stem 4-rod RFQ" should be applied for lower frequencies and "In-line stem 4-rod RFQ" should be applied for higher frequencies. The present approach is useful at an initial stage in designing of the 4-rod RFQ resonators.

## 7. ACKNOWLEDGMENTS

The author is grateful to Prof. M.Inoue, Prof. A.Noda and Dr.Y.Iwashita for their support, useful discussions about this paper and also would like to express his thanks to the members of Accelerator Laboratory, Institute for Chemical Research, Kyoto University.